\documentclass[fleqn,12pt,twoside]{article}
\usepackage{espcrc1}
\usepackage{graphicx}
\usepackage[figuresright]{rotating}

\newcommand{\AmS}{{\protect\the\textfont2
  A\kern-.1667em\lower.5ex\hbox{M}\kern-.125emS}}
\title{Identified charged particle azimuthal anisotropy
       in PHENIX at RHIC}

\author{ShinIchi Esumi,\address{Inst. of Physics, Univ. of Tsukuba,
Tenno-dai 1-1-1, Tsukuba, Ibaraki 305, Japan} for the PHENIX collaboration
\thanks{for the full PHENIX Collaboration author list and acknowledgements,
see Appendix "Collaborations" of this volume.}}
       
\begin{document}

\maketitle

\begin{abstract}
Event anisotropy is expected to have sensitivity to the early
stage of ultra-relativistic heavy-ion collisions at RHIC.
The possible formation of a quark gluon plasma (QGP) could affect 
how the initial anisotropy in the space coordinate is 
transferred into the momentum space for the final state. 
The anisotropy parameter (v$_2$) is an amplitude of the 2nd 
harmonic parameter of the azimuthal distribution with respect 
to the reaction plane.
We present here v$_2$ of identified and inclusive charged 
particles measured in the PHENIX central arm detector 
($|\eta| < 0.35$) with respect to the reaction plane defined 
at $|\eta| = 3 \sim 4$ in 200 GeV Au+Au collisions. We find 
that v$_2$ increases from central to mid-central collisions 
reaching a maximum at about 50$\%$ of the geometric cross section 
and then decreases again for more peripheral collisions. 
As a function of transverse momentum 
in minimum-bias collisions, the v$_2$ parameter increases linearly with
p$_{\rm T}$ up to p$_{\rm T}$ $\simeq$ 2 GeV/c and then saturates
for inclusive charged particles. The v$_2$ parameter of identified 
particles 
($\pi^+$, $\pi^-$, $K^+$, $K^-$, $p$ and $\overline{p}$) follow a 
hydro-dynamic behavior up to 2 GeV/c in p$_{\rm T}$, 
where the lighter mass particles have larger v$_2$ at a given
p$_{\rm T}$. However there is an indication that this trend is
reversed at around p$_{\rm T}$ $\simeq$ 2 GeV/c, where $p$ and 
$\overline{p}$ have larger v$_2$ than $\pi$ and $K$.
\end{abstract}

\section{Introduction and Experimental Setup}

The PHENIX experiment at RHIC is designed to measure a variety of 
observables in ultra-relativistic heavy-ion collisions. The primary
goal is to detect signals of the QGP. The v$_2$ measurements 
\cite{star01,phnx01} at RHIC energies have been performed 
in several experiments for inclusive charged particles and 
for identified particles. Extending the v$_2$ measurements 
at higher p$_{\rm T}$ region, especially with identified particles,  
is important to understand the origin of the v$_2$ component: hydro-dynamic 
flow of compressed matter, the production of many mini-jets, etc. It
has also been observed that v$_2$ saturates at p$_{\rm T}$ $\sim$ 2 GeV/c.
The cause of this is not yet known, but it is worth noting 
that at this momentum the particle composition is very different 
than at low momentum in that the proton yield is comparable to
the pion yield \cite{phnx02}. 

PHENIX has excellent particle identification at mid-rapidity. 
In this analysis, the time-of-flight (TOF) detector is used 
to identify final state hadrons up 
to 4 GeV/c in p$_{\rm T}$ together with the central arm tracking system 
consisting of drift- and pad-chambers for momentum reconstruction. 
The beam counters (BBC) provide collision time and z-vertex
position information. 
The two beam counters are located at $|z|$=1.5m from the collision
point, surrounding the beam pipe with 64 photo-multiplier tubes,
covering $|\eta| = 3 \sim 4$.  The large
charged multiplicity at $|\eta| = 3 \sim 4$ and non-zero signal 
of event anisotropy in this $\eta$ range enables us 
to measure the event reaction plane using the BBC with 
full azimuthal angle coverage. The standard method
\cite{e877rf,volos1} for calculating and correcting 
(flattening) the reaction plane is applied here. 
In order to obtain an improved reaction plane resolution, a 
combined reaction plane is defined by averaging the reaction 
plane angles obtained from each BBC using the elliptic moment.
The estimated resolution of the combined reaction plane,  
$<cos~ 2(\Phi_{measured}-\Phi_{truth})>$, is an average of 0.3 
with a maximum of about 0.4.
 

\section{Results and Interpretation}

Charged particles are measured in the central arm spectrometers
($|\eta| < 0.35$). 
In the present analysis, the p$_{\rm T}$ range of the particle 
identification is 0.2 $<$ p$_{\rm T}$ $<$ 3 GeV/c for pions,
0.3 $<$ p$_{\rm T}$ $<$ 2 GeV/c for kaons, and 
0.5 $<$ p$_{\rm T}$ $<$ 4 GeV/c for protons. 
The centrality of the collision is defined by the correlation 
between the total number of particles measured in the BBC and the total 
energy measured in zero degree calorimeter. 
 
\begin{figure}[h]
\begin{minipage}[t]{78mm}
\includegraphics[width=82mm]{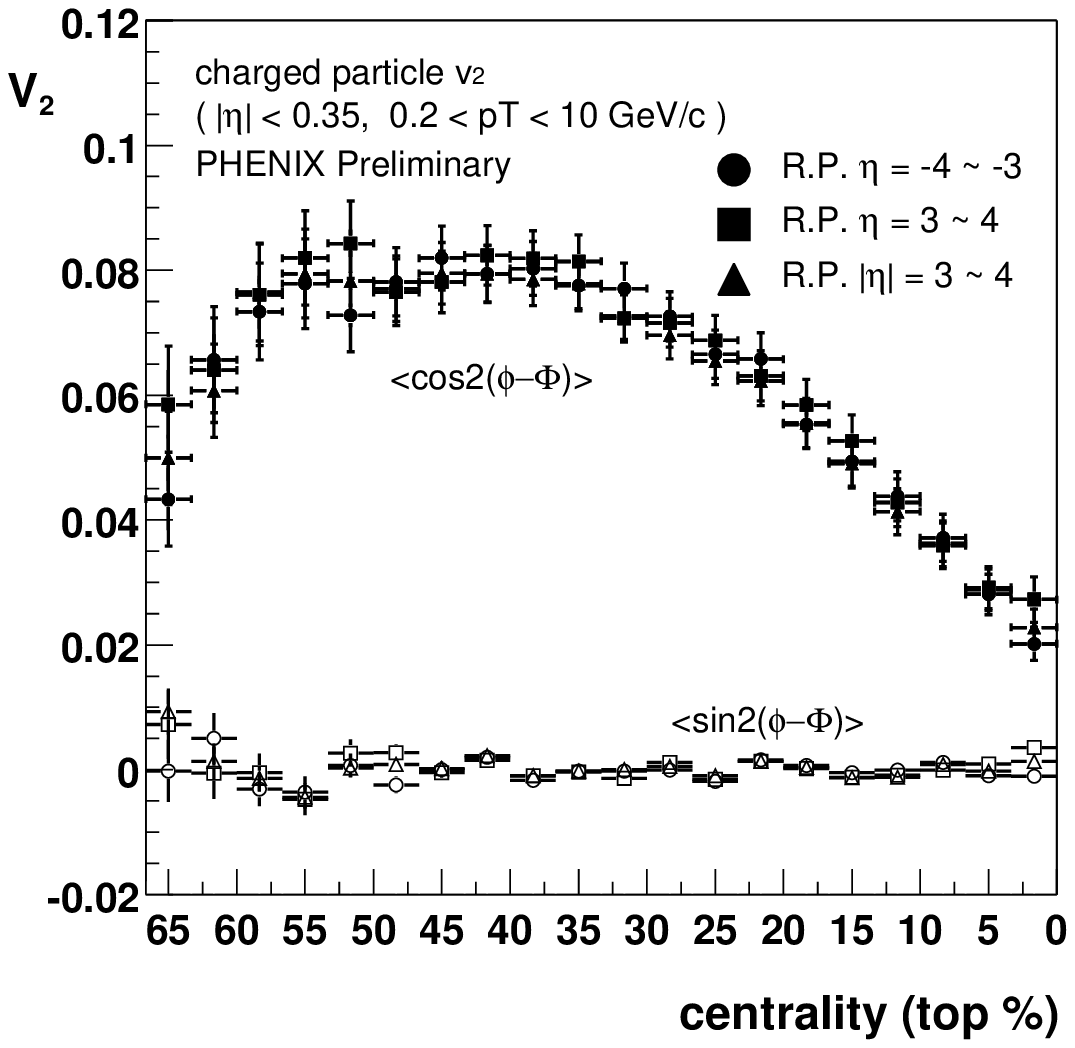}
\vskip -10mm
\caption{Centrality dependence of v$_2$ for inclusive charged particle 
measured at mid-rapidity ($|\eta| < 0.35$) and p$_{\rm T}$ 
integrated between 0.2 and 10 GeV/c with respect to the reaction 
plane defined by the two beam counters separately or combined.
Here R.P. stands for the reaction plane within the specified $\eta$ range.}
\label{qm02-esumi-fig2}
\end{minipage}
\hspace{\fill}
\begin{minipage}[t]{78mm}
\includegraphics[width=82mm]{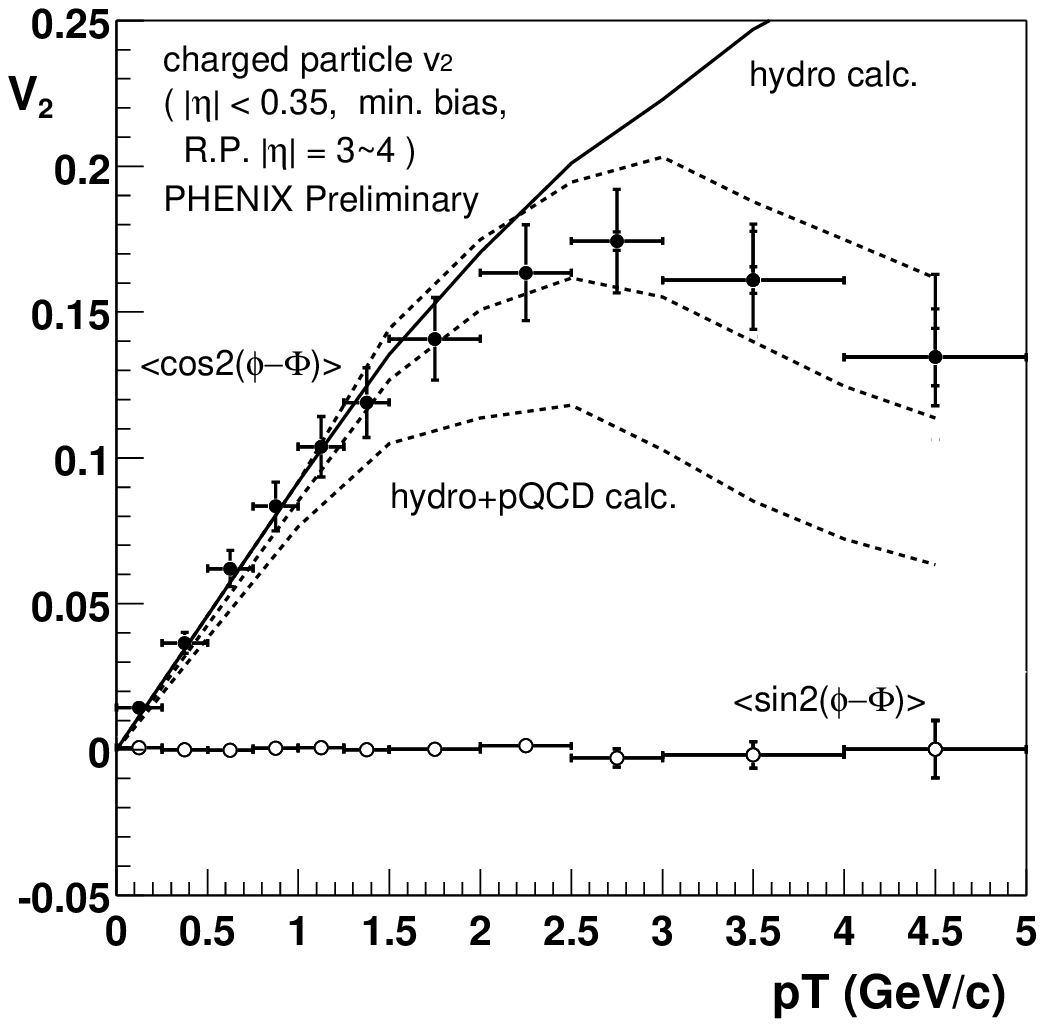}
\vskip -10mm
\caption{Transverse momentum dependence of v$_2$ for inclusive charged 
particles at mid-rapidity ($|\eta| < 0.35$) with respect to the 
combined reaction plane ($|\eta| = 3 \sim 4$) for minimum-bias 
collisions. Included are comparisons to the model calculations,
where the details are described in the text below and in refs 
\cite{houv01,gyul01}.}
\label{qm02-esumi-fig3}
\end{minipage}
\end{figure}
 
Figure \ref{qm02-esumi-fig2} shows the centrality dependence of 
v$_2$ for inclusive charged particles measured at mid-rapidity 
($|\eta| < 0.35$) with respect to three different reaction planes:
two determined using each beam counter separately, and one 
determined from the combination of the two beam counters. Here R.P. 
stands for the reaction plane within the $\eta$ range shown in the 
figure. The three results agree well, indicating that the 
resolution for the different reaction planes is well understood. 
The v$_2$ parameter decreases for both peripheral and central 
collisions with a maximum at about 50$\%$ of the geometric cross section.  
The reduction of v$_2$ for peripheral events might indicate that non-flow
contributions are less significant in this analysis because of the large
rapidity gap between the reaction plane and the central 
tracking arm acceptance of about 3 units. The data are consistent with a
$<sin~ 2(\phi-\Phi)>$ distribution (represented by the open symbols),
which is expected to be zero.
Figure \ref{qm02-esumi-fig3} shows the transverse momentum 
dependence of v$_2$ for inclusive charged particles with respect to the 
combined reaction plane for minimum-bias events. The data 
above a p$_{\rm T}$ of 2 GeV/c clearly show a deviation from 
the linear behavior seen at smaller p$_{\rm T}$. 
The solid line is from a hydro-dynamic calculation \cite{houv01}. 
The three dashed lines are from a hydro + pQCD calculation with 
three different gluon densities (from top to bottom, 
dN$^{\rm g}$/dy = 1000, 500, and 200) \cite{gyul01}.

\vskip -10mm
\begin{figure}[h]
\begin{minipage}[t]{108mm}
\includegraphics[width=115mm]{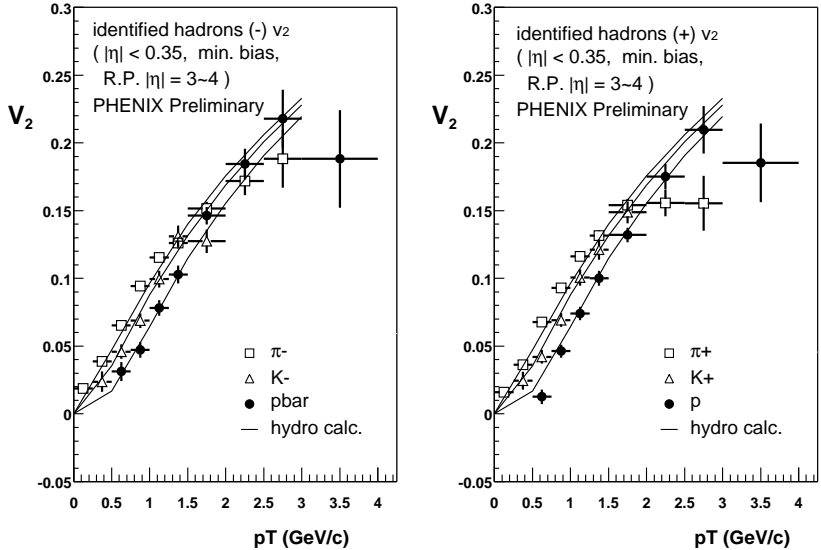}
\end{minipage}
\hspace{\fill}
\begin{minipage}[t]{48mm}
\vskip -80mm
\caption{Transverse momentum dependence of v$_2$ for identified 
particles, $\pi^-$, $K^-$, $\overline{p}$ (left) and $\pi^+$, 
$K^+$, $p$ (right). The solid lines represent a
hydro-dynamic calculation \cite{houv01} including a first-order 
phase transition with a freeze out temperature of 120 MeV 
for $\pi$, $K$ and $p$ from upper to lower curves, respectively.}
\label{qm02-esumi-fig4}
\end{minipage}
\end{figure}

\begin{figure}[h]
\begin{minipage}[t]{108mm}
\includegraphics[width=115mm]{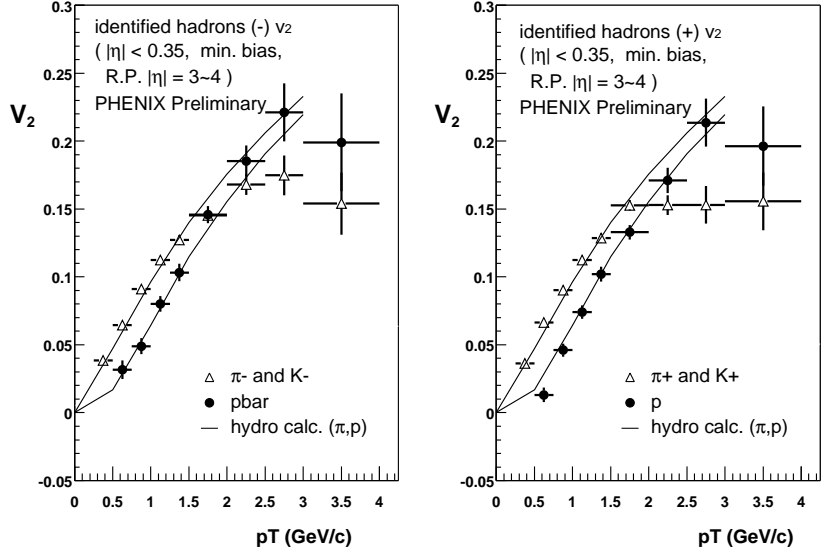}
\end{minipage}
\hspace{\fill}
\begin{minipage}[t]{48mm}
\vskip -80mm
\caption{Transverse momentum dependence of v$_2$ for  
$\pi^-$ and $K^-$ compared to $\overline{p}$ (left) and for 
$\pi^+$ and $K^+$ compared to $p$ (right). 
The solid lines are hydro-dynamic calculation 
\cite{houv01} for $\pi$ (upper curve) and $p$ (lower curve).
A clear switch over between $\pi$ + $K$ and proton is seen
at about p$_{\rm T}$ = 2 GeV/c.}
\label{qm02-esumi-fig5}
\end{minipage}
\end{figure}

\vskip -7mm
Figure \ref{qm02-esumi-fig4}, the transverse momentum 
dependence of v$_2$ for identified particles is shown. 
The solid circles show $p$ and $\overline{p}$, the open 
triangles show $K^+$ and $K^-$, and the open squares show 
$\pi^+$ and $\pi^-$. The left panel shows negatively charged 
particles, while the right panel shows positively charged particles. 
The solid lines represent hydro-dynamic calculation \cite{houv01}
including a first-order phase transition with a freeze out temperature 
of 120 MeV. The data show that at lower p$_{\rm T}$ ($<$ 2 GeV/c) 
the lighter mass particles have a larger v$_2$ at a given p$_{\rm T}$, 
which is reproduced by the model calculations. However, the data 
and the calculation deviate at higher p$_{\rm T}$. 
Lighter particles seem to deviate from the hydro-dynamic 
behavior at lower p$_{\rm T}$ than for the heavier particles. 
Consequently, the mass dependence of v$_2$ for p$_{\rm T}$
above 2 GeV/c is reversed. Since the particle identification capability 
for $K/p$ separation goes up to 4 GeV/c, the combined $\pi$ and $K$ can
be compared with protons up to 4 GeV/c. The v$_2$ value from the combined 
$\pi$ and $K$ are shown in figure \ref{qm02-esumi-fig5} in order 
to clearly show their difference from the $p$ and $\overline{p}$. 
The reversed trend of v$_2$ at p$_{\rm T}$ above 2 GeV/c is clearly seen.

\section{Summary}

The value of v$_2$ for identified and inclusive charged particles 
are measured at mid-rapidity with respect to the reaction plane defined 
in the forward and backward rapidity regions in 200 GeV Au+Au collisions 
in the PHENIX experiment at RHIC. The v$_2$ value of inclusive charged 
particles decreases for both peripheral and central collisions with 
a maximum at about 50$\%$ of the geometric cross section. 
The v$_2$ value increases linearly with p$_{\rm T}$ 
at lower p$_{\rm T}$ and saturates around 2 GeV/c.
The v$_2$ parameter for identified particles follows a hydro-dynamic 
behavior up to 2 GeV/c in p$_{\rm T}$, 
where the lighter mass particles have larger v$_2$ at a given
p$_{\rm T}$. However, there is an indication that the trend of this mass
dependence reverses at higher p$_{\rm T}$.

\end{document}